\documentclass[12pt]{article}
\usepackage{graphicx}
\begin{document}

\vspace{2.5cm}
\begin {center}
\fontsize{14}{1}\selectfont {\bf Ordering dynamics with two non-excluding options: Bilingualism in
language competition}
\end {center}
\normalsize

\bigskip

\centerline{Xavier Castell\'o, V\'{\i}ctor M. Egu\'{\i}luz, and Maxi San Miguel}

\bigskip
\centerline{IMEDEA (CSIC-UIB), Campus Universitat Illes Balears}

\centerline{E-07122 Palma de Mallorca, Spain}

\bigskip
\centerline{e-mail: xavi@imedea.uib.es}

\bigskip
{{\bf Abstract}:  We consider a modification of the voter model in
which a set of interacting elements (agents) can be in either of
two equivalent states (A or B) or in a third additional mixed AB
state. The model is motivated by studies of language competition
dynamics, where the AB state is associated with bilingualism. We
study the ordering process and associated interface and coarsening
dynamics in regular lattices and small world networks. Agents in
the AB state define the interfaces, changing the interfacial noise
driven coarsening of the voter model to curvature driven
coarsening. We argue that this change in the coarsening mechanism
is generic for perturbations of the voter model dynamics. When
interaction is through a small world network the AB agents restore
coarsening, eliminating the metastable states of the voter model.
The time to reach the absorbing state scales with system size as
$\tau \sim \ln N$ to be compared with the result $\tau \sim N$ for
the voter model in a small world network.}

\bigskip
\bigskip

\centerline{{\bf 1.INTRODUCTION}}
\bigskip

Understanding  the complex collective behavior of many particle
systems in terms of a microscopic description based on the
interaction rules among the particles is the well established
purpose of Statistical Physics. This micro-macro paradigm
\cite{Schelling} is also shared by Social Science studies based on
agent interactions (Agent Based Models). In many cases parallel
research in both disciplines goes far beyond superficial
analogies. For example, Schelling's model \cite{Schelling} of
residential segregation is mathematically equivalent to the
zero-temperature spin-exchange Kinetic Ising model with vacancies.
Cross-fertilization between these research fields opens
interesting new topics of research \cite{Granada}. In this context
the consensus problem is a general one of broad interest: the
question is to establish when the dynamics of a set of interacting
agents that can choose among several options leads to a consensus
in one of these options, or alternatively, when a state with
several coexisting social options prevails \cite{cise}. For an
equilibrium system the analogy would be with an order-disorder
transition. For nonequilibrium dynamics we rely on ideas of
studies of domain growth and coarsening in the kinetics of phase
transitions \cite{Gunton83}, where dynamics is dominated by
interface motion. Microscopic interaction rules include two
ingredients that determine the ultimate fate of the system, either
homogenous consensus state or spatial coexistence of domains of
different options. These ingredients are: i) the interaction
mechanism between particles/agents, and ii) the network of
interactions. Interactions in complex networks is a relatively
recent paradigm in statistical physics \cite{RevModPhys}. A
general still open question is the study of coarsening in complex
networks \cite{BoyerMiramontes}.

Language competition is a particular example of consensus problems
that motivates the present work. It refers to the dynamics of
language use in a multilingual social system due to individuals
interacting in a social network. Recent interest in this problem
has been triggered by the model proposed by Abrams and Strogatz
(AS-model) \cite{abrams} to account for data of extinction of
endangered languages \cite{crystal}. Other different problems of
language dynamics include those of language evolution (dynamics of
language structure) and language cognition (learning processes).
Among these, {\it semiotic dynamics}, considered in the context of
the {\it naming game} \cite{roma}, is also an example of consensus
problems. The seminal paper of Abrams and Strogatz \cite{abrams},
as well as others along the same line \cite{finland,wang,spain},
belong to the general class of mean-field population dynamics
studies based on nonlinear ordinary differential equations for the
populations of speakers of different languages. Other studies
implement microscopic agent-based-models with speakers of many or
few languages \cite{minett,schulze,kosmidis,nostre} as reviewed in
\cite{schulze2}.

The microscopic version \cite{nostre} of the AS-model for the
competition of two equivalent languages is equivalent to the voter
model
\cite{liggett,Kaprivsky96,Chate,castellanoepl,suchecki,Redner04}.
The voter model is a prototype lattice spin-model of
nonequilibrium dynamics for which $d=2$ is a critical dimension
\cite{Chate}: For regular lattices with $d>2$ coarsening does not
occur and, in the thermodynamic limit, the system does not reach
one of the homogenous absorbing states (consensus states). The
same phenomenon occurs in complex networks of interaction of
effective large dimensionality where a finite system gets trapped
in long-lived heterogeneous mestastable states
\cite{castellanoepl,suchecki,Redner04}. From the point of view of
interaction mechanisms, the voter model is one of random imitation
of a state of a neighbor. A different mechanism (for $d>1$) of
majority rule is the one implemented in a zero-temperature
spin-flip kinetic Ising (SFKI) model \cite{Krapivsky}. Detailed
comparative studies of the consequences of these two mechanisms in
different interaction networks have been recently reported
\cite{castellano}. From the point of view of coarsening and
interface dynamics, a main difference is that, in the voter model
coarsening is driven by interfacial noise, while for a SFKI
coarsening is curvature driven with surface tension reduction.

The voter and SFKI models are two-option models (spin $+1$ and
spin $-1$) with two equivalent global attractors for the system.
Kinetics of multi-option models like Potts or clock models were
addressed long ago \cite{Kaski}. More recently, a related model
proposed by Axelrod \cite{Axelrod} has been studied in some detail
\cite{CastellanoAxelrod,Klemm}. This is a multi-option model but,
in general, its nonequilibrium dynamics does not minimize a
potential leading to a thermodynamic equilibrium state like in
traditional statistical physics \cite{JEDC}. On the other hand, the kinetics
of the simplest three-options models \cite{BEG,Vazquez,Wio} has
not been studied in great detail.

We are here interested in the class of 3-state models for which
two states are equivalent (spin $\pm 1$, state A or B) and a third
one is not (spin $0$, state AB). Different dynamical microscopic
rules can be implemented for such choice of individual states,
some of which can be regarded as constrained voter-model dynamics
\cite{Vazquez}. The choice of dynamical rules in this paper is
dictated by our motivation of considering bilingual individuals in
the competition dynamics of two languages \cite{spain,minett}. We
will consider here two socially equivalent languages. The possible
state of the agents are speaking either of these languages (A or
B) or a third non-equivalent bilingual state (AB). In the context
of the consensus problem this introduces a special ingredient in
the sense that the options are not excluding: there is a possible
state of the agents (bilinguals) in which there is coexistence of
two possible options. In a more general framework, the problem
addressed here is that of competition or emergence of social norms
\cite{pujol} in the case where two norms can coexist at the individual level.

In this paper, and building upon a proposal by Minett and Wang
\cite{minett} we study a microscopic model of language competition
which reduces to the microscopic AS-model \cite{nostre} when
bilingual agents are not taken into account. Our presentation in
the remaining sections of the paper is of general nature for the
abstract problem of ordering dynamics of a modified voter model in
which a third mixed AB state is allowed. We aim to explore
possible mechanisms for the stabilization of two options
coexistence, possible metastable sates, and the role of AB states
(bilingual individuals) and interaction network (social structure)
in these processes. To this end we analyze the growth mechanisms
of A or B spatial domains (monolingual domains), the dynamics at
the interfaces (linguistic borders), and the role of AB states
(bilingual individuals) in processes of domain growth. This is
done in regular lattices and in complex networks of interaction.
Generally speaking, we find that allowing for the AB state
(bilinguals) modifies the nature and dynamics of interfaces:
agents in the AB state define thin interfaces and coarsening
processes change from voter-like dynamics to curvature driven
dynamics. We argue that this change of coarsening mechanism is
generic for perturbations of the voter model.

The outline of the paper is as follows: Section 2 describes our
microscopic model which is analyzed in a 2-dimensional regular
lattice in Section 3. In Section 4 we describe the dynamics of the model
in a small world network \cite{strogatz}. Section 5 contains a
summary of our results.

\bigskip

\bigskip
\centerline{{\bf 2. A MODEL WITH TWO NON-EXCLUDING OPTIONS }}
\bigskip

We consider a model in which an agent {\it i} sits in a node
within a network of N individuals and has $k_{i}$ neighbours. It
can be in three possible states: {\it A}, agent choosing option A (using
language A); {\it B}, agent choosing option B (using language  B); and
{\it AB}, agent in a state of coexisting options (bilingual
agent using both languages,  A and B). States A and B are
equivalent states.

The state of an agent evolves according to the following rules:
starting from a given initial condition, at each iteration we
choose one agent {\it i} at random and we compute the local
densities for each of the three communities in the neighbourhood of node {\it i},
$\sigma_{i}$ ({\it i}=A, B, AB). The agent changes state according
to the following transition probabilities proportional to the
local density of agents belonging to a community choosing a given
option ($\sigma_{A}+\sigma_{B}+\sigma_{AB}=1$)\cite{prefactor}:

$$  p_{A \rightarrow AB}=\frac{1}{2}\sigma_{B}, \quad  \quad   p_{B \rightarrow AB}=\frac{1}{2}\sigma_{A} \hspace*{0.1cm};  \eqno (1) $$
$$  p_{AB \rightarrow B}=\frac{1}{2}(1-\sigma_{A}),  \quad  \quad  p_{AB \rightarrow A}=\frac{1}{2}(1-\sigma_{B}).  \eqno (2)$$

Equation (1) gives the probabilities for an agent {\it i} to move
away from a single-option community, A or B, to the AB community. They are proportional to the density of agents in
the opposed single-option state in the neighbourhood of {\it i}.
On the other hand, equation (2) gives the probabilities for an
agent to move from the AB community towards the A or B communities.
They are proportional to the local density of agents with the
option to be adopted, including those in the AB state
($1-\sigma_{j}=\sigma_{i} + \sigma_{AB}$, {\it i, j}=A,B). It is
important to note that a change from state A to state B or vice
versa, always implies an intermediate step through the AB
state. These dynamical rules reflect the special character of
the third AB-state as one of coexisting options. They define a
modification of the two state voter model to account for the AB
state. For the voter model the transition probabilities are simply
given by $ p_{A \rightarrow B}=\sigma_{B}, \quad  p_{B \rightarrow
A}=\sigma_{A}$. These are equivalent to the adoption by the agents of the opinion of a randomly chosen neighbour. In our simulation we use random asynchronous node
update and a unit of time includes N iterations so that each node
has been updated on average once every time step.

An analysis of the mean field equations for this model shows the
existence of three fixed points: two of them stable and
equivalent, corresponding to consensus in the state A or B; and
another one unstable, with non-vanishing values for the global
densities of agents in the 3 states, A, B and AB. In order to
describe the microscopic ordering dynamics, in which we take into
account finite size effects and the topology of the network of
interactions, we use as an order parameter an ensemble average
interface density $\left<\rho\right>$. This is defined as the density of
links joining nodes in the network in different states
\cite{Chate,suchecki}. For random initial conditions $\left<\rho (t=0)\right>=2/3$.
The decrease of $\left<\rho\right>$ towards the value $\rho=0$ corresponding
to an absorbing state describes the coarsening process with growth
of spatial domains in which agents are in the same state.

\bigskip
\bigskip
\centerline{{\bf 3. COARSENING IN A REGULAR LATTICE}}
\bigskip

We first consider the dynamics on a 2-dimensional regular lattice
with four neighbours per node. We start from random initial
conditions: random spatial distribution of 1/3 of the population
in state A, 1/3 in state B and 1/3 in state AB. In Figure 1 we
show the time evolution for a typical realization: state A takes
over the system, while the opposite option B disappears. On the
average consensus in either of the two equivalent states A or B is reached with
probability $1/2$. We observe an early very fast decay of the
interface density and of the total density of agents in the state
AB, $\Sigma_{AB}$, followed by a slower decay corresponding to the coarsening dynamical stage.
This stage lasts until a finite size fluctuation
makes one of the sates A or B dominate, and the density of AB agents
disappears together with the density of agents in the option (A or
B) that vanishes.

\vspace*{1cm}
\begin{figure}[hbt]
\begin{center}
\includegraphics[angle=0,scale=0.35]{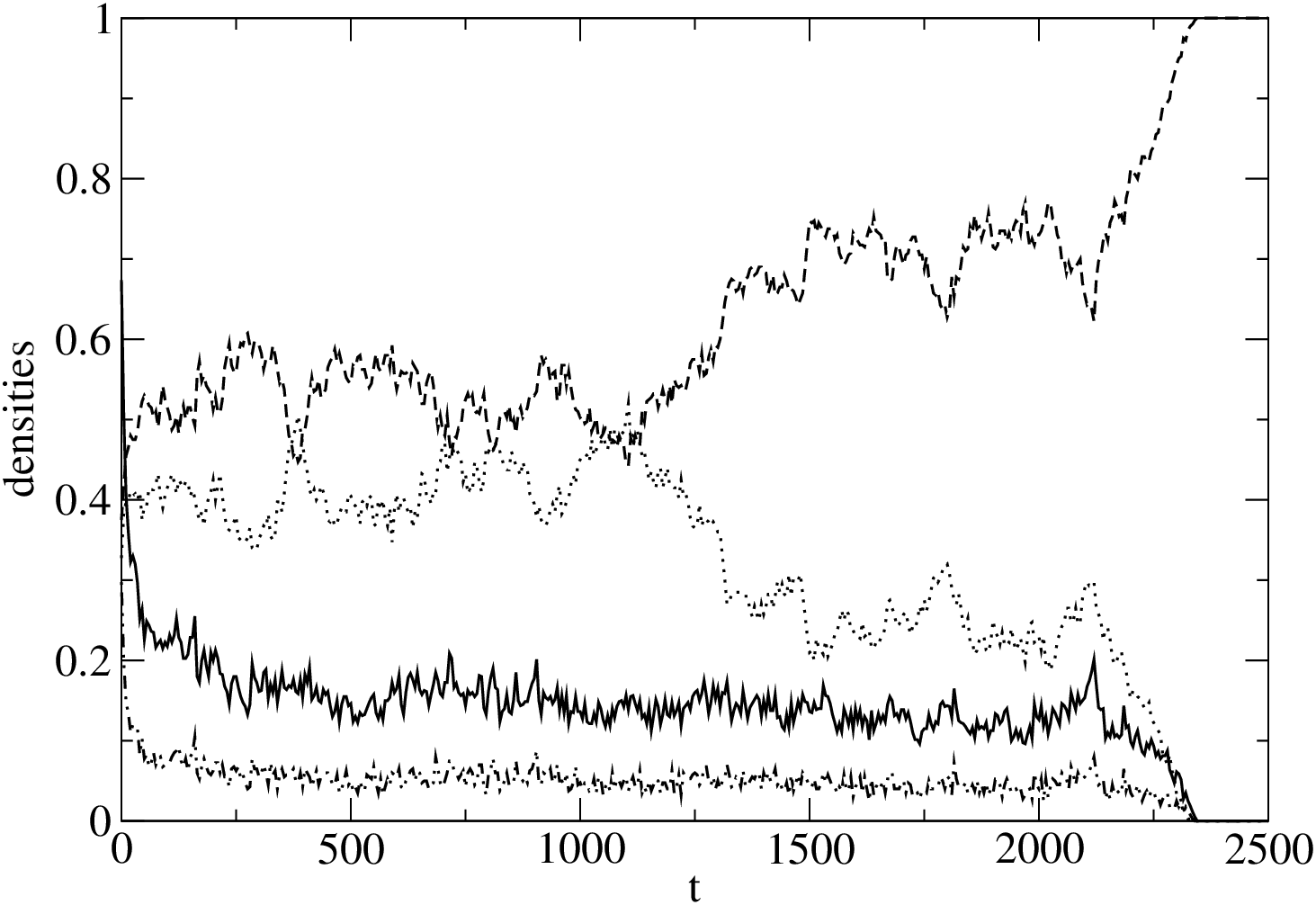}
\end{center}
\caption{Time evolution of the total densities of agents in the
three states, $\Sigma_{{\it i}}$ ({\it i}= {\it A, B, AB}),  and
the interface density, $\rho$. One realization in a population of
$N=400$ agents. From top to bottom: $\Sigma_{{\it A}}$, $\Sigma_{{\it B}}$, $\rho$, $\Sigma_{{\it AB}}$.}
\end{figure}

In Figure 2 we show the time evolution of the interface density
and of the total density of AB agents, averaged over different
realizations. For the relaxation towards one of the absorbing
states (dominance of either A or B) both the average interface
density and the average density of AB agents decay following a
power law with the same exponent, $\left<\rho\right> \sim
\left<\Sigma_{AB}\right> \sim t^{-0.45}$. This indicates that the
evolution of the average density of the AB agents is correlated
with the interface dynamics. Several systems sizes are shown in
order to see the effect of finite size fluctuations. During the
coarsening stage describe by the power law behavior, spatial
domains of the A and B community are formed and grow in size.
Eventually a finite size fluctuation occurs (as the one shown in
Figure 1) so that the whole system is taken to an absorbing state
in which there is consensus in either the A or B option. The time
scale to reach the absorbing state can be estimated to scale as
$\tau \sim N^{2}$ since at that time $\left<\rho\right> \sim 1/N$.
During the coarsening process spatial domains of AB agents are
never formed. Rather, during an early fast dynamics AB agents
place themselves in the boundaries between A and B domains. This
explains the finding that the density of AB agents follows the
same power law than the average density of interfaces. We have
also checked the intrinsic instability of an AB community: an
initial AB domain disintegrates very fast into smaller A and B
domains, with AB agents  just placed at the interfaces.  The role
of third AB state is therefore identified as a mechanism to modify
the dynamics of the interface.

\vspace*{1cm}

\begin{figure}[hbt]
\begin{center}
\includegraphics[angle=0,scale=0.35]{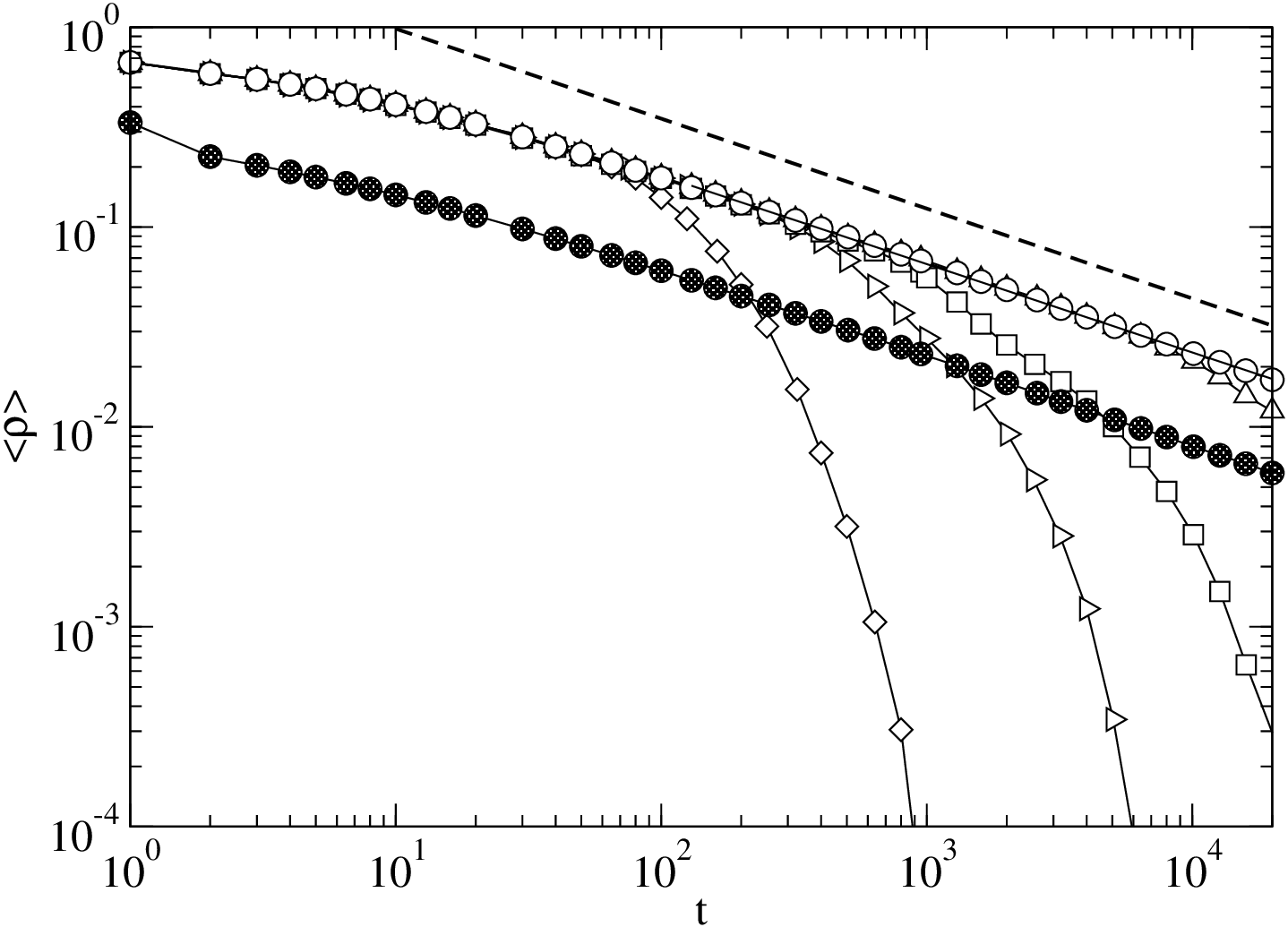}
\end{center}
\caption{ Time evolution of the averaged interface density $\left<\rho\right>$ in a 2-dimensional regular lattice for different system sizes. From left to right: $N=10^{2}$, $20^{2}$, $30^{2}$, $100^{2}$, $300^{2}$ agents (empty figures). The averaged global density of AB agents, $<\Sigma_{AB}>$, for $N=300^{2}$ agents is also shown (filled circles). Averaged over 100-1000 realizations depending on the system size. Dashed line for reference: $\left<\rho\right> \sim t^{-0.45}$}
\end{figure}

Our result for the growth law of the characteristic length of a A
or B domain is compatible with the well known exponent $0.5$
associated with domain growth driven by mean curvature and surface
tension reduction observed in SFKI models. However, systematic
deviations from the exponent $0.5$ are observed, which can be due
to non trivial logarithmic corrections. In 3-dimensional lattices,
we also find an exponent close to $0.5$ which substantiates the
claim that curvature reduction is the dominant mechanism at work.
Still, the model analyzed here is a modification of the two state
voter model for which coarsening in a $d=2$ square lattice occurs
by a different mechanism, interfacial noise, such that
$\left<\rho\right> \sim (\ln t)^{-1}$ \cite{Chate,Kaprivsky96}.
For a finite system the time to reach an absorbing state scales as
$\tau \sim N\ln(N)$ \cite{Kaprivsky92, nostre}. Therefore,
introduction of the AB state, in spite of the small number of
agents surviving in that state, implies a nontrivial modification
of the dynamics. Indeed, in our simulations we observe the
formation of well defined interfaces between A and B domains,
populated by AB agents, that evolve by a curvature driven
mechanism. On the qualitative side, the inclusion of the AB agents
gives rise to a much faster coarsening process but it also favors
a longer dynamical transient in which domains of the two competing
options coexist (larger lifetime time to reach the absorbing state
for large fixed $N$).

A natural question that these results pose is if the crossover
from interfacial noise dynamics of the voter model to curvature
driven dynamics is generic for any structural modification of the
voter model. To check this idea we have considered the coarsening
process in a 2-dimensional lattice in which agents can choose
between two excluding options (states A and B) and the dynamics is
as defined above but with transition probabilities:

$$  p_{A \rightarrow B}=\sigma_{B}-\epsilon \sin{2\pi\sigma_{B}}, \quad    p_{B \rightarrow A}=\sigma_{A}-\epsilon \sin{2\pi\sigma_{A}}, \quad  \epsilon \leq \frac{1}{2\pi}  \eqno (3) $$

The parameter $\epsilon$ measures the strength of the term that
perturbs the interaction rules of the voter model. This
perturbation of the voter model implies that the probability of
changing option is no longer a linear function of the density of
neighbouring agents in the option to be adopted. With the
perturbation term chosen here there is a nonlinear reinforcing (of
order $\epsilon$) of the effect of the local majority: the
probability to make the change $A \rightarrow B$ is larger
(smaller) than $\sigma_{B}$ when $\sigma_{B}>1/2$ ($\sigma_{B}<1/2$). In
particular, we note that for $ \epsilon \neq 0$ , the conservation
law of the ensemble average magnetization, a characteristic
symmetry of the voter model, is no longer fulfilled.
For later comparison we recall that in the zero-temperature SFKI
the local majority determines, with probability one, the change of
option: $p_{A \rightarrow B} = 1 (0)$ if $\sigma_{B}>1/2 (\sigma_{B}<1/2)$.

Our results for the exponent $x$ in a power law fitting
$\left<\rho\right> \sim t^{-x}$ for the modified voter model
defined by eq. (3) \cite{extension_SFKI} are shown in Fig. 3 for different values of
$\epsilon$ \cite{8_veins}. For very small values of $\epsilon$ we observe an
exponent $x \sim 0.1$ compatible with the logarithmic decay
($\left<\rho\right> \sim (\ln t)^{-1}$) of the voter model, as
obtained in \cite{nostre}. However, for small, but significant
values of $\epsilon$ there is a change to a value $x \sim 0.5$
associated with curvature driven coarsening.

We conclude that a small arbitrary perturbation of the transition probabilities of the voter model dynamics leads to a new interface dynamics, equivalent to the one found in Section 2 by including a third state where options are non-excluding. This indicates that voter model dynamics is very sensitive to perturbations of its dynamical rules.

\vspace*{1cm}
\begin{figure}[hbt]
\begin{center}
\includegraphics[angle=0,scale=0.35]{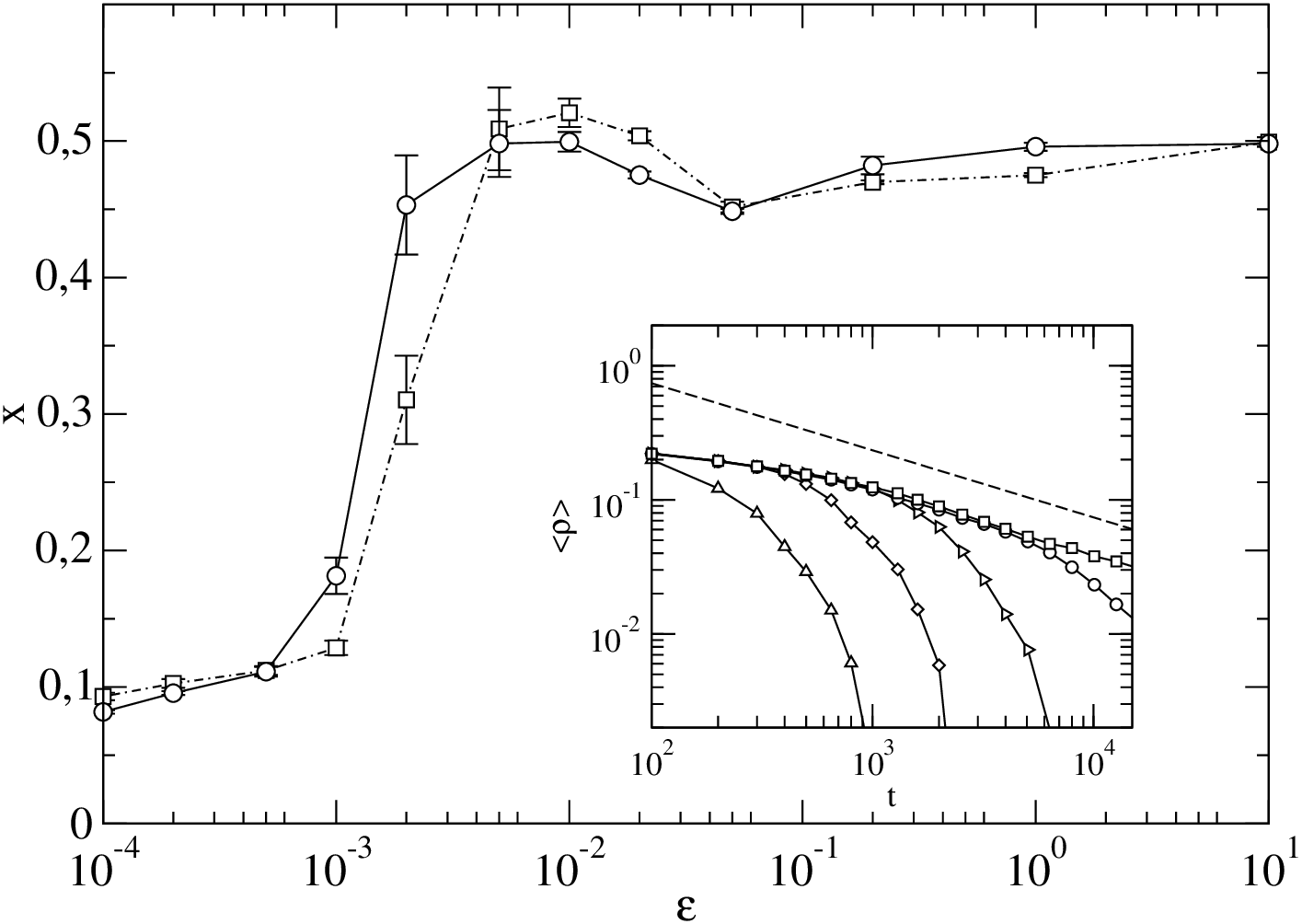}
\end{center}
\caption{ Characteristic coarsening exponent $x$
($\left<\rho\right> \sim t^{-x}$) for the modified voter model
(eq.3) \cite{extension_SFKI} as a funtion of the perturbation parameter $\epsilon$. From left to right, 
systems of sizes $N=400^{2}$ (averaged over 50 realizations), $300^{2}$ (averaged over 75 realizations). Inset: time evolution of the average interface densitiy, for $\epsilon=0.01$. From left to right: $N=20^{2}$, $50^{2}$, $100^{2}$, $200^{2}$, $400^{2}$ agents. Given a value of $\epsilon$, for large enough system sizes a power law for the average interface density decay is found. Dashed line for reference: $\left<\rho\right> \sim t^{-0.5}$}
\end{figure}

\bigskip
\centerline{{\bf 4. COARSENING IN A SMALL WORLD NETWORK}}
\bigskip

We next consider the dynamics of the model defined in Sect.2 on a
small world network constructed following the algorithm of Watts
\& Strogatz \cite{strogatz}: starting from a two dimensional
regular lattice with four neighbours per node, we rewire with probability {\it p} each of the
links at random, getting in this way a partially disordered
network with long range interactions throughout it.

\vspace*{1cm}

\begin{figure}[hbt]
\begin{center}
\includegraphics[angle=0,scale=0.35]{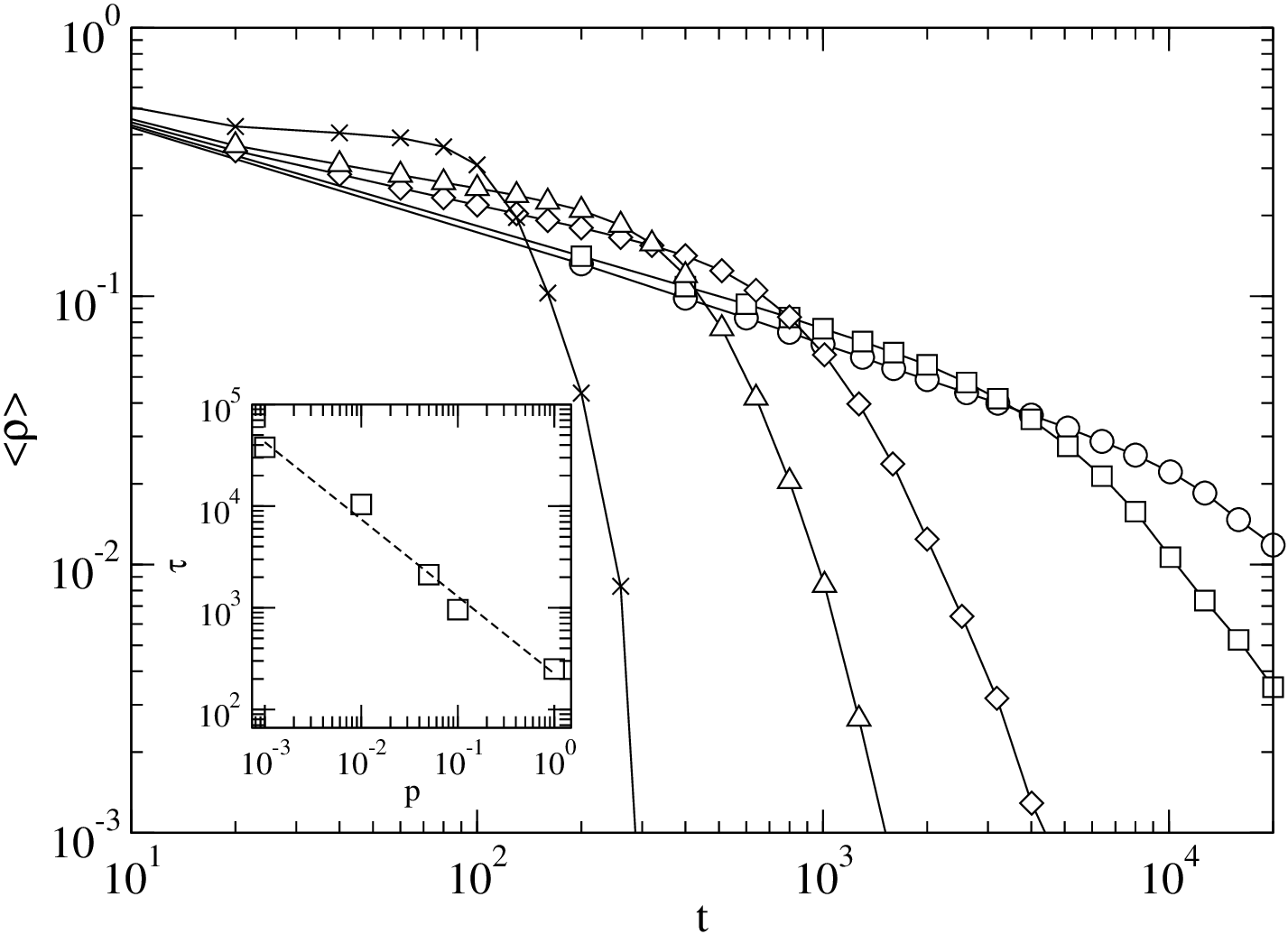}
\end{center}
\caption{ Time evolution of the average interface density
$\left<\rho\right>$ in small world networks with different values
of the rewiring parameter $p$. From left to right: p=1.0, 0.1, 0.05, 0.01, 0.0. For comparison the case $p=0$ for a
regular network and the case $p=1$ corresponding to a random
network are also shown. The inset shows the dependence of the
characteristic lifetime $\tau$ with the rewiring parameter $p$. The dashed line corresponds to the power law fit $\tau \sim p^{-0.76}$.
Population of $100^{2}$ agents, averaged over 500 realizations.}
\end{figure}

In Figure 4 we show the evolution of the average interface density
for different values of $p$. As for the regular lattice we also
observe here a dynamical stage of coarsening with a power law
decrease of $\left<\rho\right>$ followed by a fast decay to the A
or B absorbing caused by a finite size fluctuation. During the
dynamical stage of coarsening, the A and B communities have
similar size, while the total density of AB agents is much smaller. In the range of
intermediate values of $p$ properly corresponding to a small world
network, increasing the rewiring parameter $p$ has two main
effects: i) the coarsening process is notably slower; ii) the
characteristic time of the dynamics $\tau$, which we define as the
time when $\left<\rho\right>$ sinks below a given small value,
drops following a power law (inset of Figure 4): $\tau \sim
p^{-0.76}$, so that the absorbing state is reached much faster as
the network becomes disordered.

\vspace*{1cm}
\begin{figure}[hbt]
\begin{center}
\includegraphics[angle=0,scale=0.35]{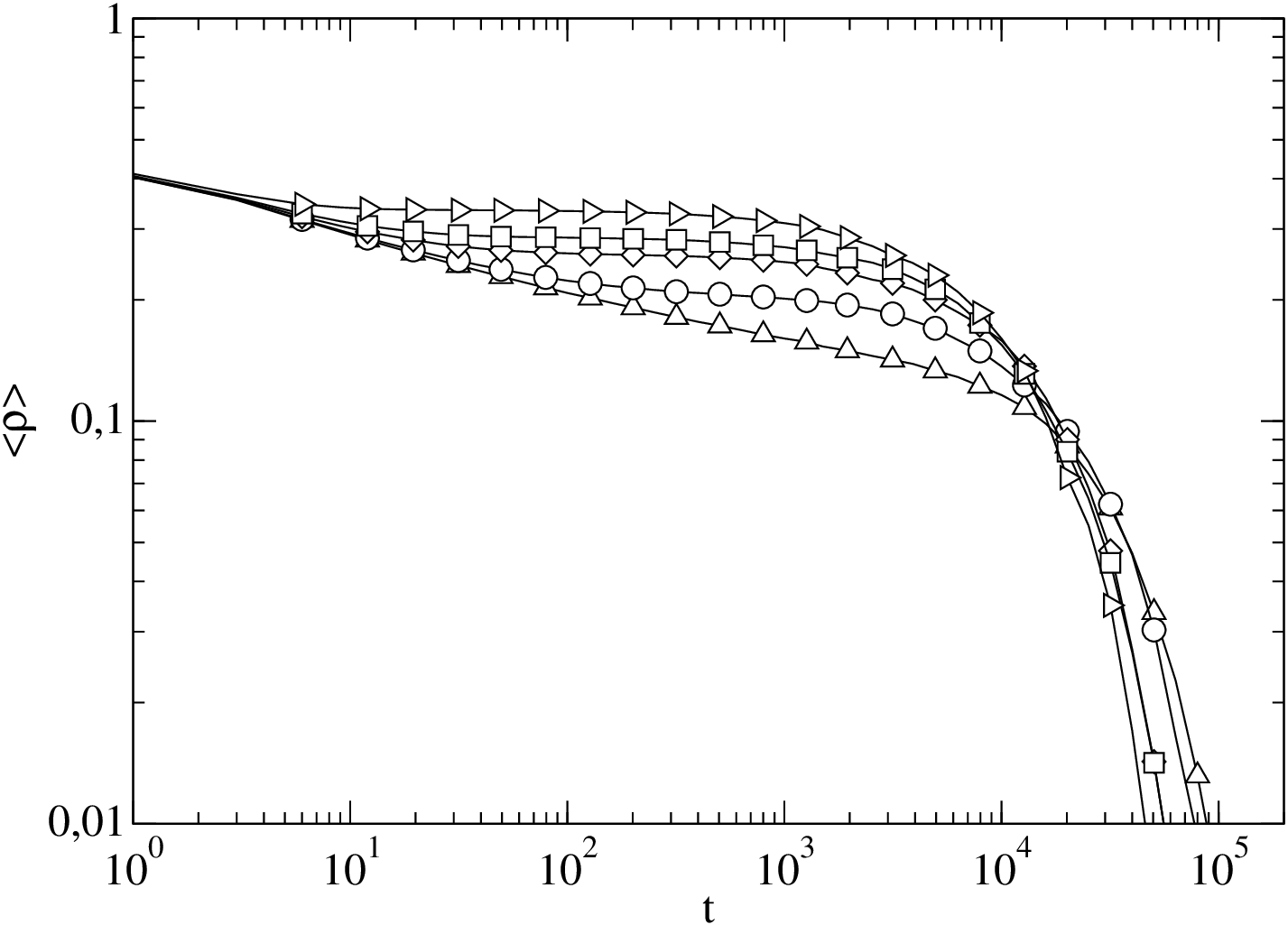}
\end{center}
\caption{Time evolution of the average interface density
$\left<\rho\right>$ for the voter model in a small world network
with different values of $p$. From up to bottom, p=1.0, 0.1, 0.05, 0.01, 0.0. Population of $100^{2}$ agents,
averaged over 900 realizations. }
\end{figure}

To understand the role of the AB state in the ordering dynamics in
a small world network, the results of Fig. 4 should be compared
with the ones in Fig. 5 for the two state voter model
in the same small world network \cite{footnotecastellano}. In
contrast with the model with two non-excluding options (Section 2), moderate
values of $p$ stop the coarsening process leading to dynamical
metastable states characterized by a plateau regime for the
average interface density \cite{castellanoepl,suchecki}. However
the lifetime of these states is not very sensitive to the value of
$p$,  with the characteristic time of the dynamics being just
slightly smaller than the one obtained in a regular lattice
($p=0$). This is a different effect than the strong dependence on
$p$ found for these characteristic times when AB agents are
included in the dynamics. Comparing the results of Figs. 4 and 5
for a fixed intermediate value of $p$, we observe that including
AB agents in the dynamics on a small world network of interactions
allows the coarsening process to take place, and it also produces
an earlier decay to the absorbing state.

\vspace*{1cm}
\begin{figure}[hbt]
\begin{center}
\includegraphics[angle=0,scale=0.35]{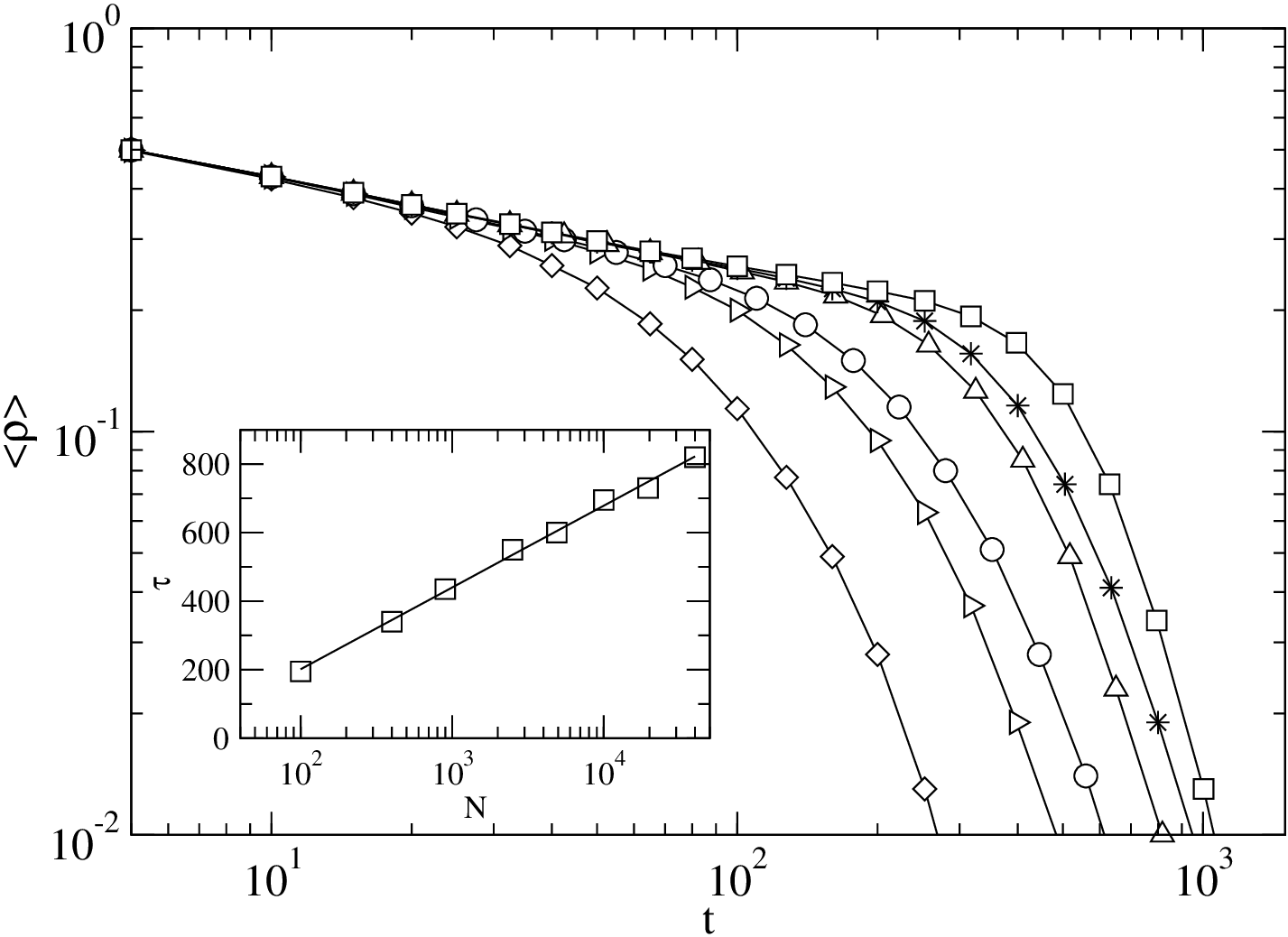}
\end{center}
\caption{ Time evolution of the averaged interface density,
$\left<\rho\right>$, for different values of the population size
($N$) in a small world network with $p=0.1$. $N=10^{2}$, $20^{2}$,
$30^{2}$, $70^{2}$, $100^{2}$, $200^{2}$ from left to right.
Averaged over 1000 realizations in 10 different networks. Inset:
dependence of the characteristic time $\tau$ (time when
$\left<\rho\right>$ sinks below a given small value; 0.03 in this
figure) with the system size: $\tau \sim \ln(N)$.}
\end{figure}

System size dependence for a fixed value of the rewiring parameter
$p$ is analyzed in Figure 6. We observe that the initial stage of
coarsening process is grossly independent of system size, but the
characteristic time of the dynamics scales with the system size
$N$ as $\tau \sim \ln(N)$. For the two state voter model $\tau
\sim N$ \cite{suchecki}. Therefore the faster decay to the
absorbing state caused by the AB agents in a large system
interacting through a small world network is measured by the ratio
$ \frac{\tau_{AB}} {\tau_{voter}} \arrowvert_{SW} \sim \frac{\ln(N)}{N} $. This faster decay is the inverse than the one found
for the regular lattice where the same ratio of time scales is $
\frac{\tau_{AB}} {\tau_{voter}} \arrowvert_{Lattice} \sim \frac{N^2}{N \ln(N)}
$.

\bigskip
\bigskip
\centerline{{\bf 5. SUMMARY AND CONCLUSIONS}}
\bigskip

We have studied the nonequilibirum transient dynamics of approach
to the absorbing state for a modified voter model defined in Sect.
2 in which the interacting agents can be in either of two
equivalent states (A or B) or in a third mixed state (AB). A
global consensus state (A or B) is reached with probability one.
A domain of agents in the AB state is not
stable and the density of AB-agents becomes very small after an
initial fast transient. In spite of these facts, the AB-agents
produce an essential modification of the processes of coarsening
and domain growth, changing the interfacial noise dynamics of the
voter model into a curvature driven interface dynamics
characteristic of two-option models based on local majorities updating rules.
We have argued that this effect is generic for small structural
modifications of the random imitation dynamics of the voter model.
We have also considered the effect of the topology of the network
of interactions studying our dynamical model in a small world
network. While for the original voter model the small world
topology results in long lived metastable states in which
coarsening has become to a halt \cite{castellanoepl,suchecki}, the
AB-agents restore the processes of coarsening and domain growth.
Additionally, they speed-up the decay to the absorbing state by a
finite size fluctuation. We obtain a characteristic time that
scales with system size as $\tau \sim \ln N$ to be compared with
the result $\tau \sim N$ for the voter model.

From the point of view of recent studies of linguistic dynamics
the modified voter model allowing for two non-exluding options, is an extension of the microscopic
version \cite{nostre} of the Abrams-Strogatz model \cite{abrams} for two socially equivalent
languages, to include the effects of
bilingualism (AB-agents) \cite{minett} and social structure. Within the assumptions and limitations of our
model, our results imply that bilingualism and small world social structure
are not efficient mechanisms to stabilize language diversity. On
the contrary they are found to ease the approach to absorbing
monolingual states by an obvious effect of smoothing the
communication across linguistic borders.

We acknowledge financial support form the MEC(Spain) through
project CONOCE2 (FIS2004-00953). X.C. also acknowledges financial support from a phD fellowship of the Conselleria d'Economia, Hisenda i Innovaci\'o del Govern de les Illes Balears.


\begin{thebibliography}{99}
\bibitem{prefactor}The prefactor $1/2$ comes from the original model \cite{minett}, when taking
into account non-equivalent options A and B. Here it can be
interpreted as an inertia that limits the maximum probability for
changing sate to $1/2$ setting a microscopic time scale.
\bibitem{extension_SFKI} It is possible to extend the perturbation
of the voter model defined by eq. (3) letting $\epsilon$ to take
any positive value, and choosing a modified transition
probabilitiy defined as follows: $p_{A\rightarrow B}$ as given by
eq. (3) for values of  $\epsilon$ such that $0\leq p_{A\rightarrow
B} \leq 1$; $p_{A\rightarrow B}=0(1)$ for values of $\epsilon$
such that eq. (3) gives $p_{A\rightarrow B}<0(>1)$. The limit
$\epsilon \rightarrow \infty$, corresponds to the step-function
transition probability of the SFKI model at T=0.
\bibitem{8_veins} In these simulations we have taken a 2-dimensional lattice
with eight neighbours per node so that more possible values are
allowed for the perturbation term in eq. (3).
\bibitem{footnotecastellano} Note that the small world network considered in
\cite{castellanoepl} is obtained by a rewiring process of a $d=1$
regular lattice.
\end{thebibliography}
\end{document}